\documentclass[11pt]{article}
\usepackage{graphicx}

\newcommand{\pas}[1]{\textbf{\sf #1}}
\newcommand{\ttb}[1]{\texttt{\textbf{#1}}}
\newcommand{\tbi}[1]{\textbf{\tit{#1}}}

\newcommand{\tit}[1]{\textit{#1}}
\newcommand{\ttt}[1]{\texttt{#1}}
\newcommand{\res}[1]{\textbf{#1}}

\newcommand{\inputpic}[3]{
\begin{figure}[!t]
  \begin{center}
  \includegraphics[width=#2]{pics/#1.eps}
  \end{center}
  \caption{\emph{#3}}
  \label{#1}
\end{figure}
}

\begin{document}

\title{\textbf{An $O(m)$ Algorithm for\\ Cores Decomposition of Networks}}

\author{
   Vladimir Batagelj, Matja\v{z} Zaver\v{s}nik \\
   University of Ljubljana, FMF, Department of Mathematics,\\
   and IMFM Ljubljana, Department of TCS, \\
   Jadranska ulica 19, 1\,000 Ljubljana, Slovenia\smallskip \\
   \texttt{vladimir.batagelj@uni-lj.si}\\
   \texttt{matjaz.zaversnik@fmf.uni-lj.si}
}

\date{ }

\maketitle

\begin{abstract}
   The structure of large networks can be revealed by partitioning them to
   smaller parts, which are easier to handle. One of such decompositions is
   based on $k$--cores, proposed in 1983 by Seidman. In the paper an efficient,
   $O(m)$, $m$ is the number of lines, algorithm for determining the cores
   decomposition of a given network is presented.
\end{abstract}


\section{Introduction}

``One of the major concerns of social network analysis is identification
of cohesive subgroups of actors within a network. Cohesive subgroups are
subsets of actors among whom there are relatively strong, direct, intense,
frequent, or positive ties'' (\cite{WaF94}, p. 249). Several notions were
introduced to formally describe cohesive groups: cliques, $n$--cliques,
$n$--clans, $n$--clubs, $k$--plexes, $k$--cores, lambda sets, \ldots
For most of them it turns out that they are algorithmically difficult
(NP hard \cite{GJ} or at least quadratic), but for cores a very efficient
algorithm exists. We describe it in details in this paper.


\section{Cores}

The notion of core was introduced by Seidman in 1983 \cite{SEIDMAN}.

Let $G=(V,L)$ be a graph. $V$ is the set of \tit{vertices} and $L$ is the
set of \tit{lines} (\tit{edges} or \tit{arcs}). We will denote $n = |V|$ and
$m = |L|$. A subgraph $H=(W,L|W)$ induced by the set $W$ is a $k$-\emph{core}
or a \emph{core of order} $k$ iff $\forall v\in W: \deg_H(v)\geq k$ and $H$
is a maximum subgraph with this property. The core of maximum order is also
called the \emph{main} core. The \emph{core number} of vertex $v$ is the
highest order of a core that contains this vertex.

The degree $\deg(v)$ can be: in-degree, out-degree, in-degree $+$ out-degree,
\ldots dete\-rmining different types of cores.

In figure \ref{cores} an example of cores decomposition of a given graph is
presented. From this figure we can see the following properties of cores:

\inputpic{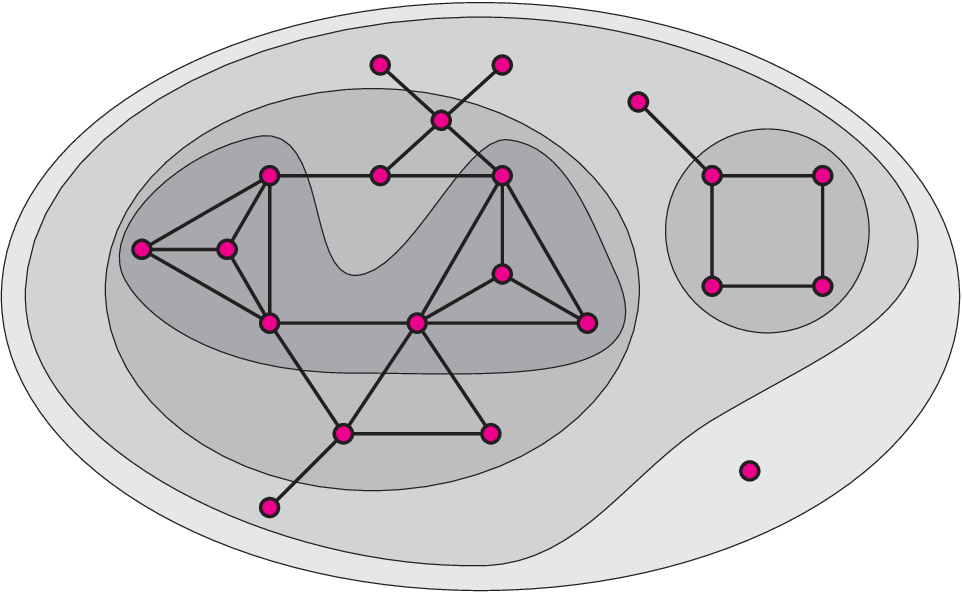}{80mm}{0, 1, 2 and 3 core}

\begin{itemize}
\item The cores are nested: \ $i<j \ \ \Longrightarrow \ \ H_j\subseteq H_i$
\item Cores are not necessarily connected subgraphs.
\end{itemize}


\section{Algorithm}

Our algorithm for determining the cores hierarchy is based on the following
property \cite{BMCON}:

\begin{quote}
If from a given graph $G=(V,L)$ we recursively delete all vertices, and lines
incident with them, of degree less than $k$, the remaining graph is the $k$-core.
\end{quote}
The outline of the algorithm is as follows:

\begin{tabbing}
xxxxx\=xxx\=xxx\=xxx\=xxx\=xxx\= \kill \\
INPUT: graph $G=(V,L)$ represented by lists of neighbors \\
OUTPUT: table $core$ with core number for each vertex \\
\\
1.1\>compute the degrees of vertices; \\
1.2\>order the set of vertices $V$ in increasing order of their degrees;\\
2\>\pas{for each} $v\in V$ in the order \pas{do begin}\\
2.1\>\>$core\lbrack v\rbrack := degree\lbrack v\rbrack$; \\
2.2\>\>\pas{for each} $u \in Neighbors(v)$ \pas{do}\\
2.2.1\>\>\>\pas{if} $degree\lbrack u\rbrack > degree\lbrack v\rbrack$ \pas{then begin}\\
2.2.1.1\>\>\>\>$degree\lbrack u\rbrack := degree\lbrack u\rbrack - 1$;\\
2.2.1.2\>\>\>\>reorder $V$ accordingly\\
\>\>\>\pas{end}\\
\>\pas{end};
\end{tabbing}
In the refinements of the algorithm we have to provide efficient
implementations of steps 1.2 and 2.2.1.2.


\section{Detailed Algorithm}

We describe an implementation of the algorithm in a Pascal like language.

Structure \ttt{graph} is used to represent a given graph $G=(V,L)$. We will not
describe the structure into details, because there are several possibilities,
how to do this. We assume that the vertices of $G$ are numbered from 1 to $n$.
The user has also to provide functions \ttt{size} and \ttt{in Neighbors},
described in the table:

\begin{table}[!ht]
\centering
\begin{tabular}{@{\hspace{0mm}}ll@{\hspace{0mm}}}
\hline
\ttt{name($parameters$)} & returned value \\
\hline
\ttt{size($G$)} & number of vertices in graph $G$ \\
\ttt{$u$ in Neighbors($G$,$v$)} & $u$ is a not yet visited neighbor of vertex $v$ in graph $G$ \\
\hline
\end{tabular}
\end{table}

Using an adequate representation of graph $G$ (lists of neighbors) we can
implement both functions to run in constant time.

\inputpic{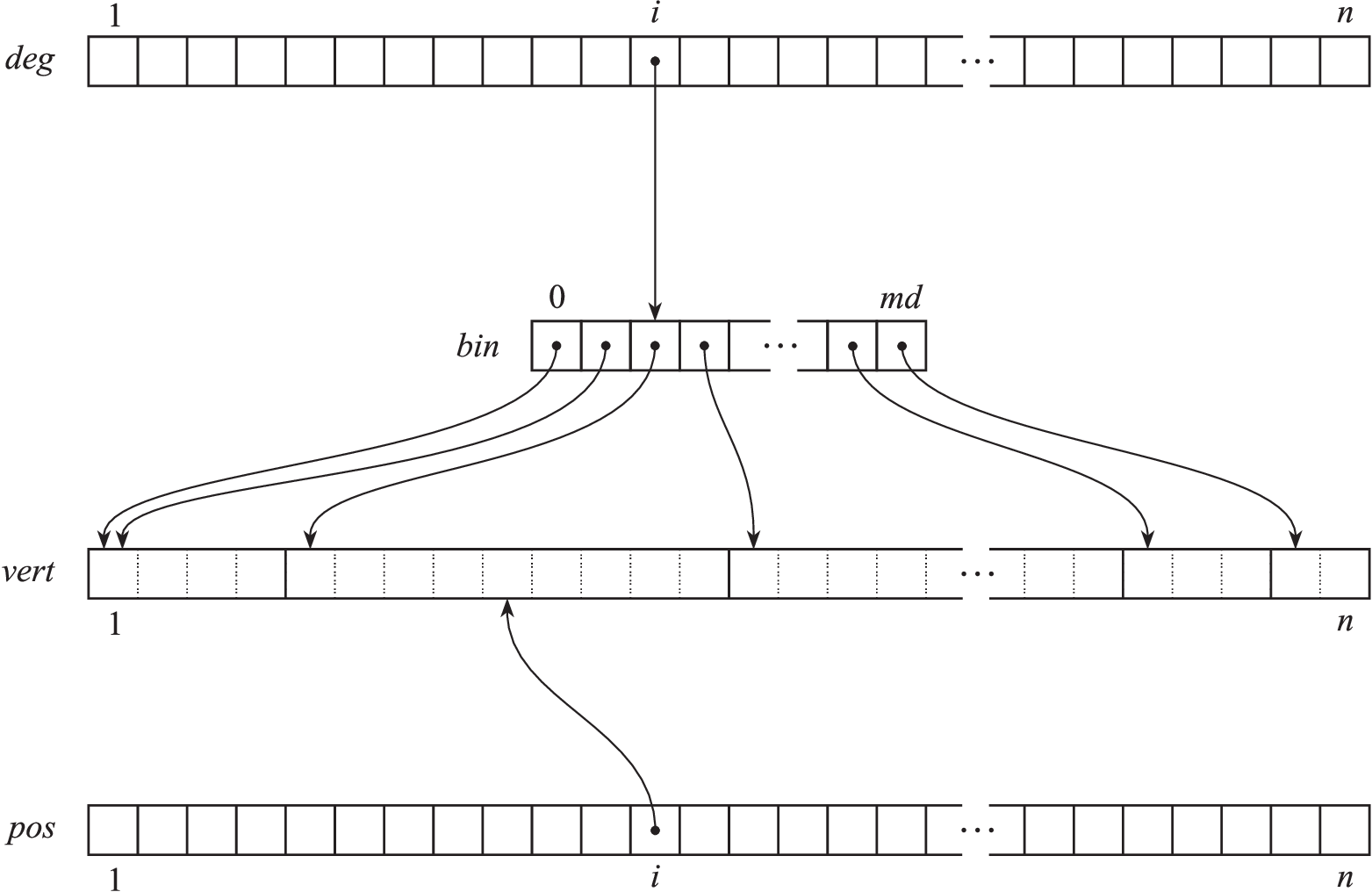}{12cm}{Arrays}

Two types of integer arrays (\ttt{tableVert} and \ttt{tableDeg}) are also
introduced. Both of them must be of length at least $n$. The only difference
is how we index their elements. We start with index 1 in \ttt{tableVert}
and with index 0 in \ttt{tableDeg}.

The algorithm is implemented by procedure \ttt{cores}. The input is graph $G$,
represented by variable \ttt{g} of type \ttt{graph}, the output is array \ttt{deg}
of type \ttt{tableVert} containing core number for each vertex of graph $G$.

\begin{table}
\renewcommand{\tablename}{Algorithm}
\caption{\emph{The Cores Algorithm for Simple Undirected Graphs}}
\medskip
\renewcommand{\baselinestretch}{0.8}\normalsize
\begin{ttfamily}
01\ \ \res{procedure}\ cores(\res{var}\ g:\ graph;\ \res{var}\ deg:\ tableVert);\\
02\ \ \res{var}\\
03\ \ \ \ \ n,\ d,\ md,\ i,\ start,\ num:\ integer;\\
04\ \ \ \ \ v,\ u,\ w,\ du,\ pu,\ pw:\ integer;\\
05\ \ \ \ \ vert,\ pos:\ tableVert;\\
06\ \ \ \ \ bin:\ tableDeg;\\
07\ \ \res{begin}\\
08\ \ \ \ \ n\ :=\ size(g);\ \ md\ :=\ 0;\\
09\ \ \ \ \ \res{for}\ v\ :=\ 1\ \res{to}\ n\ \res{do}\ \res{begin}\\
10\ \ \ \ \ \ \ \ d\ :=\ 0;\ \ \res{for}\ u\ in\ Neighbors(g,\ v)\ \res{do}\ inc(d);\\
11\ \ \ \ \ \ \ \ deg[v]\ :=\ d;\ \ \res{if}\ d\ >\ md\ \res{then}\ md\ :=\ d;\\
12\ \ \ \ \ \res{end};\\
13\ \ \ \ \ \res{for}\ d\ :=\ 0\ \res{to}\ md\ \res{do}\ bin[d]\ :=\ 0;\\
14\ \ \ \ \ \res{for}\ v\ :=\ 1\ \res{to}\ n\ \res{do}\ inc(bin[deg[v]]);\\
15\ \ \ \ \ start\ :=\ 1;\\
16\ \ \ \ \ \res{for}\ d\ :=\ 0\ \res{to}\ md\ \res{do}\ \res{begin}\\
17\ \ \ \ \ \ \ \ num\ :=\ bin[d];\\
18\ \ \ \ \ \ \ \ bin[d]\ :=\ start;\\
19\ \ \ \ \ \ \ \ inc(start,\ num);\\
20\ \ \ \ \ \res{end};\\
21\ \ \ \ \ \res{for}\ v\ :=\ 1\ \res{to}\ n\ \res{do}\ \res{begin}\\
22\ \ \ \ \ \ \ \ pos[v]\ :=\ bin[deg[v]];\\
23\ \ \ \ \ \ \ \ vert[pos[v]]\ :=\ v;\\
24\ \ \ \ \ \ \ \ inc(bin[deg[v]]);\\
25\ \ \ \ \ \res{end};\\
26\ \ \ \ \ \res{for}\ d\ :=\ md\ \res{downto}\ 1\ \res{do}\ bin[d]\ :=\ bin[d-1];\\
27\ \ \ \ \ bin[0]\ :=\ 1;\\
28\ \ \ \ \ \res{for}\ i\ :=\ 1\ \res{to}\ n\ \res{do}\ \res{begin}\\
29\ \ \ \ \ \ \ \ v\ :=\ vert[i];\\
30\ \ \ \ \ \ \ \ \res{for}\ u\ in\ Neighbors(g,\ v)\ \res{do}\ \res{begin}\\
31\ \ \ \ \ \ \ \ \ \ \ \res{if}\ deg[u]\ >\ deg[v]\ \res{then}\ \res{begin}\\
32\ \ \ \ \ \ \ \ \ \ \ \ \ \ du\ :=\ deg[u];\ \ \ pu\ :=\ pos[u];\\
33\ \ \ \ \ \ \ \ \ \ \ \ \ \ pw\ :=\ bin[du];\ \ w\ :=\ vert[pw];\\
34\ \ \ \ \ \ \ \ \ \ \ \ \ \ \res{if}\ u\ <>\ w\ \res{then}\ \res{begin}\\
35\ \ \ \ \ \ \ \ \ \ \ \ \ \ \ \ \ pos[u]\ :=\ pw;\ \ vert[pu]\ :=\ w;\\
36\ \ \ \ \ \ \ \ \ \ \ \ \ \ \ \ \ pos[w]\ :=\ pu;\ \ vert[pw]\ :=\ u;\\
37\ \ \ \ \ \ \ \ \ \ \ \ \ \ \res{end};\\
38\ \ \ \ \ \ \ \ \ \ \ \ \ \ inc(bin[du]);\ \ dec(deg[u]);\\
39\ \ \ \ \ \ \ \ \ \ \ \res{end};\\
40\ \ \ \ \ \ \ \ \res{end};\\
41\ \ \ \ \ \res{end};\\
42\ \ \res{end};
\end{ttfamily}
\end{table}

We need (03-06) some integer variables and three additional arrays. Array
\ttt{vert} contains the set of vertices, sorted by their degrees. Positions
of vertices in array \ttt{vert} are stored in array \ttt{pos}. Array \ttt{bin}
contains for each possible degree the position of the first vertex of that
degree in array \ttt{vert}. See also Figure \ref{arrays} in which a Pascal
implementation of our algorithm for the case of simple undirected graph
$G=(V,E)$, $E$ is the set of edges, is presented.

In a real implementation of the proposed algorithm dynamically allocated
arrays should be used. To simplify our description of the algorithm we
replaced them by static.

At the beginning we have to initialize some local variables and arrays
(08-12). First we determine \ttt{n}, the number of vertices of graph \ttt{g}.
Then \tbi{we compute degree for each vertex} \ttt{v} \tbi{in graph}
\ttt{g} and store it into array \ttt{deg}. Simultaneously we also
compute the maximum degree \ttt{md}.

After that we \tbi{sort the vertices in increasing order of their degrees}
using bin-sort (13-25). First we count (13-14) how many vertices will be
in each bin (bin consists of vertices with the same degree). Bins are
numbered from 0 to \ttt{md}.

From bin sizes we can determine (15-20) starting positions of bins in array
\ttt{vert}. Bin 0 starts at position 1, while other bins start at position,
equal to the sum of starting position and size of the previous bin. To avoid
additional array we used the same array (\ttt{bin}) to store starting positions
of bins. Now we can put (21-25) vertices of graph $G$ into array \ttt{vert}.
For each vertex we know to which bin it belongs and what is the starting
position of that bin. So we can put vertex to the proper place, remember its
position in table \ttt{pos}, and increase the starting position of the bin we
used. The vertices are now sorted by their degrees.

In the final step of initialization phase we have to \tbi{recover starting
positions of the bins} (26-27). We increased them several times in previous
step, when we put vertices into corresponding bins. It is obvious, that the
changed starting position is the original starting position of the next bin.
To restore the right starting positions we have to move the values in array
\ttt{bin} for one position to the right. We also have to reset starting
position of bin 0 to value 1.

The \tbi{cores decomposition}, implementing the \pas{for each} loop from the
algorithm described in section 3, is done in the main loop (28-41) that runs
over all vertices \ttt{v} of graph \ttt{g} in the order, determined by table
\ttt{vert}. The core number of current vertex \ttt{v} is the current degree
of that vertex. This number is already stored in table \ttt{deg}. For each
neighbor \ttt{u} of vertex \ttt{v} with higher degree we have to decrease its
degree and move it for one bin to the left. Moving vertex \ttt{u} for one bin
to the left is operation, which can be done in constant time. First we have to
swap vertex \ttt{u} and the first vertex in the same bin. In array \ttt{pos} we
also have to swap their positions. Finally we increase starting position of
the bin (we increase previous and reduce current bin for one element).


\subsection{Time complexity}

We shall show that the described algorithm runs in time $O(\max(m,n))$.

To compute (08-12) the degrees of all vertices we need time $O(\max(m,n))$ since
we have to consider each line at most twice. The \tit{bin sort} (13-27) consists
of five loops of size at most $n$ with constant time $O(1)$ bodies -- therefore
it runs in time $O(n)$.

The statement (29) requires a constant time and therefore contributes $O(n)$ to
the algorithm. The conditional statement (31-39) also runs in constant time.
Since it is executed for each edge of $G$ at most twice the contribution of
(30-40) in all repetitions of (28-41) is $O(\max(m,n))$.

Summing up --- the total time complexity of the algorithm is $O(\max(m,n))$.
Note that in a connected network $m \geq n-1$ and therefore $O(\max(m,n)) = O(m)$.


\subsection{Adaption of the algorithm for directed graphs}

For directed simple graphs without loops only few changes in the implementation
of the algorithm are needed depending on the interpretation of the \emph{degree}.
In the case of in-degree and out-degree the function \ttt{in Neighbors} must return
next not yet visited in-neighbor and out-neighbor respectively. If degree is
defined as in-degree $+$ out-degree, the maximum degree can be at most $2n-2$.
In this case we must provide enough space for table \ttt{bin} ($2n-1$ elements).
Function \ttt{in Neighbors} must return next not yet visited in-neighbor or
out-neighbor.


\section{Example}

We applied the described algorithm for cores decomposition on a network based
on the Knuth's English dictionary \cite{DICK}.
This network has 52652 vertices (English words having 2 to 8 characters)
and 89038 edges (two vertices are adjacent, if we can get one word from another
by changing, removing or inserting a letter). The obtained network is sparse:
density is 0.0000642. The program took on PC only 0.01 sec to compute the core numbers.
In the table below the summary results are presented.

\begin{table}[!ht]
\centering
\begin{tabular}{|r|rr|rr|}
\hline
&
\multicolumn{2}{|c|}{vertices with core number $k$} &
\multicolumn{2}{|c|}{size of $k$-core} \\
$k$ & \# & \% & \# & \% \\
\hline
25 & 26 & 0.049 & 26 & 0.049 \\
16 & 34 & 0.065 & 60 & 0.114 \\
15 & 16 & 0.030 & 76 & 0.144 \\
14 & 59 & 0.112 & 135 & 0.257 \\
13 & 82 & 0.156 & 217 & 0.412 \\
12 & 200 & 0.380 & 417 & 0.792 \\
11 & 202 & 0.384 & 619 & 1.176 \\
10 & 465 & 0.883 & 1084 & 2.059 \\
9 & 504 & 0.957 & 1588 & 3.016 \\
8 & 923 & 1.753 & 2511 & 4.769 \\
7 & 1114 & 2.116 & 3625 & 6.885 \\
6 & 1590 & 3.020 & 5215 & 9.905 \\
5 & 2423 & 4.602 & 7638 & 14.507 \\
4 & 3859 & 7.329 & 11497 & 21.836 \\
3 & 5900 & 11.206 & 17397 & 33.042 \\
2 & 8391 & 15.937 & 25788 & 48.978 \\
1 & 13539 & 25.714 & 39327 & 74.693 \\
0 & 13325 & 25.308 & 52652 & 100.000 \\
\hline
\end{tabular}
\end{table}

Vertices with core number 0 are isolated vertices. Vertices with core number
1 have only one neighbor in the network. The 25-core (main core) consists of
26 vertices, where each vertex has at least 25 neighbors inside the core
(obviously this is a clique). The corresponding words are \ttt{a's}, \ttt{b's},
\ttt{c's}, \ldots, \ttt{y's}, \ttt{z's}.

The 16-core has additional 34 vertices (\ttt{an}, \ttt{on}, \ttt{ban}, \ttt{bon},
\ttt{can}, \ttt{con}, \ttt{Dan}, \ttt{don}, \ttt{eon}, \ttt{fan}, \ttt{gon},
\ttt{Han}, \ttt{hon}, \ttt{Ian}, \ttt{ion}, \ttt{Jan}, \ttt{Jon}, \ttt{man},
\ttt{Nan}, \ttt{non}, \ttt{pan}, \ttt{pon}, \ttt{ran}, \ttt{Ron}, \ttt{San},
\ttt{son}, \ttt{tan}, \ttt{ton}, \ttt{van}, \ttt{von}, \ttt{wan}, \ttt{won},
\ttt{yon}, \ttt{Zan}). There are no edges between vertices with
core number 25 and vertices with core number 16. The adjacency matrix of
the subgraph induced by these 34 vertices is presented on figure \ref{core16}.
In this matrix we can see two 17-cliques and some additional edges.

\inputpic{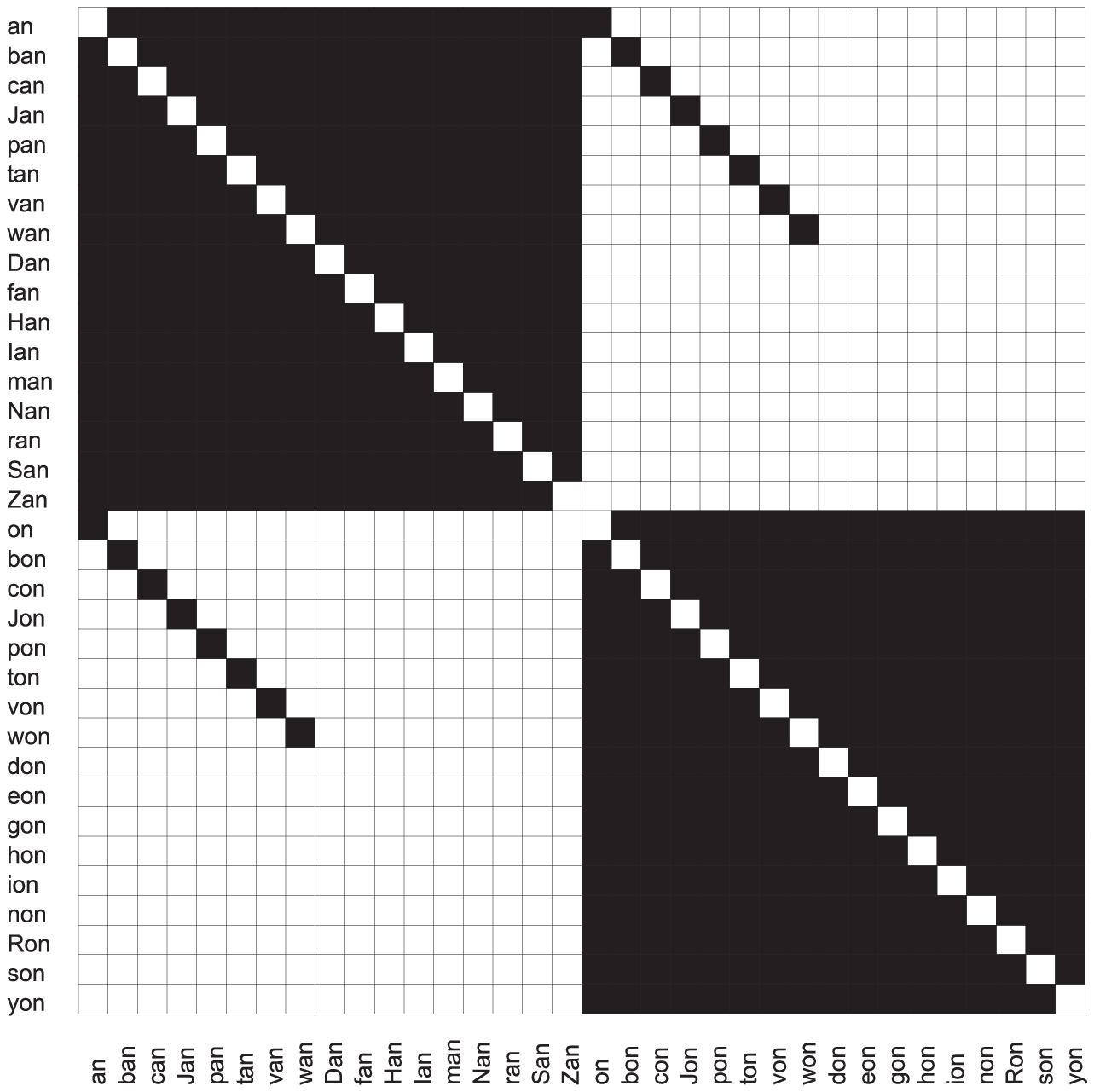}{100mm}{Adjacency matrix of 16-core without 25-core}

The 15-core has additional 16 vertices (\ttt{ow}, \ttt{bow}, \ttt{cow}, \ttt{Dow},
\ttt{how}, \ttt{jow}, \ttt{low}, \ttt{mow}, \ttt{now}, \ttt{pow}, \ttt{row},
\ttt{sow}, \ttt{tow}, \ttt{vow}, \ttt{wow}, \ttt{yow}). This is a clique again,
because only the first letters of the words are different.


\section{Conclusion}

The cores, because they can be efficiently determined, are one among few concepts
that provide us with meaningful decompositions of large networks. We expect that
different approaches to the analysis of large networks can be built on this basis.
For example, the sequence of vertices in sequential coloring can be determined
by their core numbers (combined with their degrees). Cores can also be used to
reveal interesting subnetworks in large networks \cite{BMZ,ERDOS}.

The described algorithm is implemented in program for large networks analysis
\ttb{Pajek} (Slovene word for Spider) for Windows (32 bit) \cite{BMCON}.
It is freely available, for noncommercial use, at its homepage:

\ttb{http://vlado.fmf.uni-lj.si/pub/networks/pajek/}


\section*{Acknowledgment}

This work was supported by the Ministry of Education, Science and Sport of
Slovenia, Project J1-8532. It is a detailed version of the part of the talk presented at
\emph{Recent Trends in Graph Theory, Algebraic Combinatorics, and Graph Algorithms},
September 24--27, 2001, Bled, Slovenia,


\end{document}